# Effect of Driver Behavior on Spatiotemporal Congested Traffic Patterns at Highway Bottlenecks in the Framework of Three-Phase Traffic Theory

## Boris S. Kerner*

*Daimler AG, GR/PTF, HPC: G021, 71059 Sindelfingen, Germany*

---

### Abstract

We present results of numerical simulations of the effect of driver behavior on spatiotemporal congested traffic patterns that result from traffic breakdown at an on-ramp bottleneck. The simulations are made with the Kerner-Klenov stochastic traffic flow model in the framework of three-phase traffic theory. Different diagrams of congested patterns at the bottleneck associated with different driver behavioral characteristics are found and compared each other. An adaptive cruise control (ACC) in the framework of three-phase traffic theory introduced by the author (called a "driver alike ACC" (DA-ACC)) is discussed. The effect of DA-ACC-vehicles on traffic flow, in which without the DA-ACC-vehicles traffic congestion occurs at the bottleneck, is numerically studied. We show that DA-ACC-vehicles improve traffic flow considerably without any reduction in driving comfort. It is found that there is a critical percentage of DA-ACC-vehicles in traffic flow: If the percentage of the DA-ACC-vehicle exceeds the critical one no traffic breakdown occurs at the bottleneck. A criticism of a recent "criticism of three-phase traffic theory" is presented.

---

## 1. Introduction

The understanding of traffic congestion is the key for development of many other fields of transportation science and engineering. For this reason, a huge number of publications, reviews, and books are devoted to empirical studies of traffic congestion and associated traffic flow theories (see references in Haight, 1963; Drew, 1968; Prigogine and Herman, 1971; Whitham, 1974; Wiedemann, 1974; Cremer, 1979; Newell, 1982; Papageorgiou, 1983; Leutzbach, 1988; May, 1990; Daganzo, 1997; Brackstone and McDonald, 1998; Highway Capacity Manual, 2000; Chowdhury et al., 2000; Gartner et al. (eds), 2001; Helbing, 2001; Gazis, 2002; Nagatani, 2002; Nagel et al., 2003; Mahnke et al., 2005; Rakha et al., 2009). In empirical observations, traffic breakdown in free flow occurs mostly at bottlenecks associated with, e.g., on- and off-ramps. In congested traffic, moving jams are observed. A moving jam is a localized structure of great vehicle density, spatially limited by two jam fronts; the jam propagates upstream; within the jam vehicle speed is very low. Theoretical studies of traffic congestion at bottlenecks have been made mostly either in the framework of the Lighthill-Whitham-Richards (LWR) kinematic (shock) wave traffic flow theory or in the framework of the General Motors (GM) approach by Herman, Gazis, Montroll, Potts, and Rothery.

Although there are many important achievements in the understanding of traffic congestion made in these and many other works, however, these approaches cannot explain empirical features of traffic breakdown at highway bottlenecks as found in real measured traffic data. A detailed criticism of the description of traffic breakdown and resulting congested traffic patterns within the frameworks of the LWR- and GM-approaches as well as other traffic flow theories reviewed in (May, 1990; Brackstone and McDonald, 1998; Chowdhury et al., 2000; Gartner et al. (eds), 2001; Helbing, 2001; Gazis, 2002; Nagatani, 2002; Nagel et al., 2003; Mahnke et al., 2005; Rakha et al., 2009) has been made in (Kerner, 2009). In this article, we have to limit a consideration of our critical responses *only* on "criticism of three-phase traffic theory" published in "Transportation Science" and "Transportation Research B" (Schönhof and Helbing, 2007, 2009). Readers can find all other our criticisms on earlier traffic flow theories and models in Chapter 10 of the book (Kerner, 2009).

---

\* Boris S. Kerner. Tel.: + 49-7031-4389566.
*E-mail address*: boris.kerner@daimler.com.

To describe traffic breakdown in accordance with real measured data, the author introduced three-phase traffic theory reviewed in (Kerner, 2004a, 2009). In this theory, there are (i) the free flow, (ii) synchronized flow, and (iii) wide moving jam traffic phases. The synchronized flow and wide moving jam phases associated with congested traffic are defined via the empirical definitions [S] and [J], respectively. A wide moving jam is a moving traffic jam, i.e., a localized structure of great vehicle density and low speed, spatially limited by two jam fronts, which exhibit the characteristic jam feature [J] to propagate through bottlenecks while maintaining the mean velocity of the downstream jam front. Synchronized flow [S] is defined as congested traffic that does not exhibit the jam feature [J]; in particular, the downstream front of synchronized flow is often fixed at the bottleneck.

The fundamental hypothesis of three-phase traffic theory is that steady states of synchronized flow cover a 2D-region (S) in the flow-density plane and there is a speed gap between this region and free flow states (F) at each given density (Fig. 1 (a)). This means also that within the 2D-region each driver tends to adapt its speed to the preceding vehicle ("speed adaptation") without caring what a precise space (time) gap is as long as the gap is safe (Fig. 1 (b)): There is no relationship between the flow rate and density (no fundamental diagram) even for steady states of synchronized flow. The upper boundary of the 2D-region in Fig. 1 (b) is associated with a so-called synchronization space gap denoted by $G$ that determines a synchronization time headway; the lower boundary of the 2D-region in Fig. 1 (b) is associated with a safe space gap denoted by $g_{safe}$ that determines a safe time headway.

The first microscopic traffic flow model in which many hypotheses of three-phase traffic theory have been incorporated is the Kerner-Klenov stochastic three-phase traffic flow model (Sect. 16.3 of Kerner, 2004a). The 2D-region of steady states of synchronized flow in this model is shown in Fig. 1 (c, d). This model can show traffic breakdown at a bottleneck and resulting congested patterns as found in empirical data (Kerner, 2004a, 2009). For this reason, we will use the model for all simulations presented in this article. Author's three-phase traffic theory is the theoretical basis for many new traffic flow models and control methods developed recently (see e.g., Davis, 2004, 2006, 2010; Jiang, et al., 2004; Lee et al., 2004; Jia et al., 2009; Gao et al., 2007, 2009).

In the Kerner-Klenov model, a variety of driver behavioral characteristics have been used, which influence considerably on spatiotemporal congested traffic patterns occurring at a bottleneck (Kerner, 2004a, 2009). However, in this article we limit a consideration of the influence on the patterns of two driver behaviors only:

    (i)       the speed adaptation within 2D-region of synchronized flow states of three-phase traffic theory (Kerner, 2004a) and

    (ii)      the well-known over-deceleration effect of the GM model class associated with a delay in driver deceleration that is caused by a driver reaction time (Herman, et al., 1958; Gazis et al., 1961).

To demonstrate the importance of this driver behavioral analysis for transportation engineering, we show that if some of the vehicles in traffic flow move in accordance with an adaptive cruise control (ACC) in the framework of three-phase traffic theory introduced in (Kerner, 2004b) whose main feature is the speed adaptation within 2D-region of synchronized flow states (Fig. 1 (b)), then both traffic breakdown and moving jam emergence can be prevented *without any reduction in comfortable driving*.

Before we present these novel results, in section "Background" we repeat the Kerner-Klenov model used for all simulations, basic theoretical features of traffic breakdown and resulting congested patterns at a on-ramp bottleneck as well as those driver behavioral assumption of the model that are required for the paper understanding.

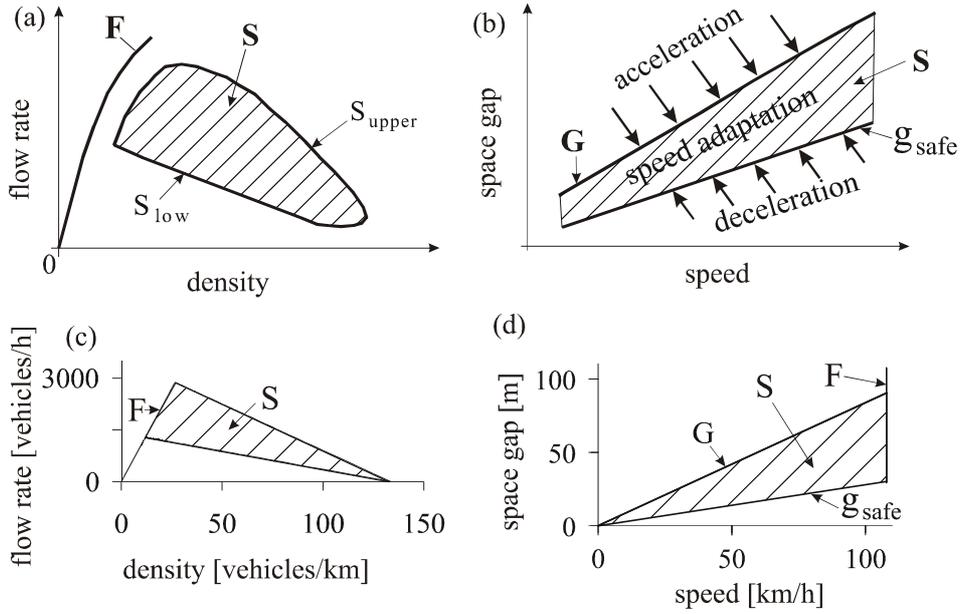

Fig. 1. Fundamental hypothesis of three-phase traffic theory (Kerner, 2004a): (a, b) Qualitative representation of the hypothesis in the flow-density (a) and speed-space gap planes (in (b) only a part of 2D synchronized flow states of (a) is shown). (b, c) Steady states of the Kerner-Klenov stochastic three-phase traffic flow model used in simulations.

## 2. Background

### 2.1. Two Classes of Homogeneous Synchronized Flow (HSF) in Three-Phase Traffic Theory

For the understanding of the paper results, firstly we should remember some general features of 2D steady states of synchronized flow associated with the fundamental hypothesis of three-phase traffic theory (dashed region in Figs. 1 and 2). Steady states of synchronized flow are homogeneous in space and time. Therefore, synchronized flow associated with these states can be called "homogeneous synchronized flow" (HSF). Thus in three-phase traffic theory, at each vehicle density and speed at which synchronized flows exists (2D dashed regions in Fig. 2), a synchronized flow can be an HSF.

In three-phase traffic theory, an HSF exhibits qualitatively different features depending of whether the HSF is associated with a state below or above the line *J* in the flow-density plane (line *J* in Fig. 2). Recall that the line *J* represents the steadily propagation of the downstream front of a wide moving jam in the flow-density plane, i.e., the slope of the line *J* is equal to the mean velocity $v_g$ of this jam front:

$$v_g = -1 / (\rho_{max} \tau_{del, jam}^{(a)}),$$

where $\rho_{max}$ is the jam density, $\tau_{\mathrm{del, \, jam}}^{(a)}$ is the mean driver time delays in acceleration at the downstream jam front. The left and right coordinates of the line $J$ are associated with the maximum jam outflow $q_{out}$ and jam density $\rho_{max}$, respectively.

The feature of HSF is as follows (Kerner, 1998): The line $J$ divides all HSF into two different HSF classes with respect to wide moving jam formation:

    (i)    stable HSF,

    (ii)    metastable HSF.

Stable HSF are associated with states below the line $J$, whereas metastable HSF are associated with states on and above the line $J$ (Fig. 4). In a stable HSF, no wide moving jams can emerge and persist independent of whether there are great fluctuations of time headways, speed, density, and/or flow rate in the HSF or not.

In contrast with the stable HSF, in a metastable HSF a wide moving jam(s) emerges *only* if speed (space gap) fluctuations appear that are greater than some critical ones. Otherwise, no wide moving jams emerge.

The general feature of metastable HSFs is as follows (Kerner, 2004a): At the same vehicle speed, the closer time headway to a safety time headway, the greater the probability of wide moving jam emergence within a metastable HSF during the same time interval.

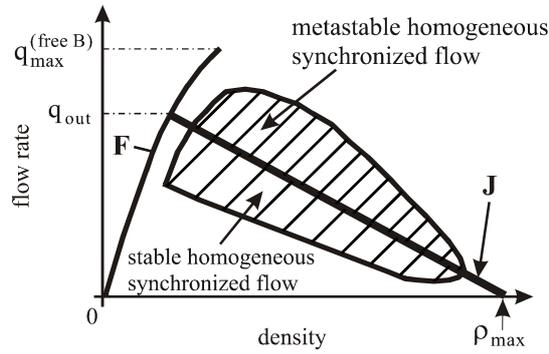

Fig. 2. Homogeneous synchronized flow (HSF) of three-phase traffic theory (Kerner, 1998). Free flow states (F) ands 2D-dashed region of synchronized flow states are taken from Fig. 1 (a).

## 2.2. Discrete Version of Stochastic Three-Phase Traffic Flow Model

For a numerical study of traffic breakdown at highway bottlenecks and resulting congested traffic on multi-lane roads we use a *discrete* (both in time and space) version (Kerner and Klenov, 2009) of the Kerner-Klenov stochastic continuum in space three-phase traffic flow model in which vehicles move on a two-lane road (Sect. 18.2 of Kerner, 2004a): Rather than the continuum space co-ordinate, a discretized space co-ordinate with a small enough discretization cell $\delta x$ is used. Consequently, the speed and acceleration (deceleration) discretization intervals are

$$\delta v = \delta x / \tau \quad \text{and} \quad \delta a = \delta v / \tau, \qquad (1)$$

respectively, where model time step $\tau = 1$ s. In comparison with the continuum in space model, the discrete model version allows us a more accurate study of phase transitions in synchronized flow (Kerner and Klenov, 2009). In this paper, we limit consideration of the model in which all drivers and vehicles are identical ones.

It must be noted that the Kerner-Klenov model incorporates both three-phase traffic theory (Kerner, 1998, 1999, 2004a) and many pertinent ideas about simulations of a variety of driver time delays in different driving situations of earlier traffic flow models in the framework of the fundamental diagram hypothesis introduced by Herman, Montroll, Potts, Rothery, Gazis (Herman, et al., 1959; Gazis, et al., 1961), Newell (1961), Nagel, Schreckenberg, Schadschneider, and co-workers (Nagel and Schreckenberg, 1992; Barlovic', et al., 1998), Bando, Sugiyama, and colleagues (Bando, et al., 1995) and many other groups.

In the discrete model version, the vehicle coordinate, speed, and acceleration (deceleration) are dimensionless integer values measured respectively in values $\delta x$, $\delta v$, and $\delta a$. To emphasize the physical sense of model formulae, we remain in the formulae time step $\tau$ explicitly; however, $\tau$ in all formulae below should be considered the dimensionless value $\tau = 1$. The model is as follows (Sect. 18.2 of Kerner, 2004a; Kerner and Klenov, 2009):

$$v_{n+1} = \max(0, \min(v_{free}, \tilde{v}_{n+1} + \xi_n, v_n + a\tau, v_{s,n})), \quad x_{n+1} = x_n + v_{n+1}\tau, \qquad (2)$$

$$\tilde{v}_{n+1} = \max(0, \min(v_{free}, v_{s,n}, v_{c,n})), \qquad (3)$$

$$v_{c,n} = \begin{cases} v_n + \Delta_n & \text{for} \quad g_n \leq G_n \\ v_n + a_n\tau & \text{for} \quad g_n > G_n \end{cases}, \qquad (4)$$

$$\Delta_n = \max(-b_n\tau, \min(a_n\tau, \ v_{\ell,n} - v_n)), \qquad (5)$$

where index n corresponds to the discrete time $t = n\tau$, $n = 0,1,2,...$; $v_n$ is the vehicle speed at time step $n$; $x_n$ is the vehicle co-ordinate; $\tilde{v}_n$ is the speed calculated without a noise component $\xi_n$; $g_n = x_{\ell,n} - x_n - d$ is the space gap (space headway) between vehicles, $d$ is the vehicle length, the lower index $\ell$ marks functions (or values) related to the preceding vehicle; $v_{free}$ is the maximum speed in free flow (line F in Fig. 1(c)); $v_{s,n}$ is a safe speed (see below); $a_n \geq 0$ and $b_n \geq 0$; $a$ is the maximum acceleration; $G_n$ is a synchronization space gap:

$$G_n = G(v_n, v_{\ell,n}), \qquad (6)$$

$$G(u, w) = \max(0, \lfloor k\tau u + a^{-1}\phi_0 u(u - w) \rfloor), \qquad (7)$$

where k>1, $\phi_0 > 0$ are constants, $\lfloor z \rfloor$ denotes the integer part of a real value z.

The safe speed is

$$v_{s,n} = \min(v_n^{(safe)}, v_\ell^{(a)} + \frac{g_n}{\tau}), \qquad (8)$$

where

$$v_\ell^{(a)} = \max\left(0, \min\left(v_{\ell,n}^{(safe)}, v_{\ell,n}, g_{\ell,n}/\tau\right) - a\tau\right), \qquad (9)$$

is an anticipation speed, $v_n^{(safe)} = \lfloor v^{(safe)}(g_{n,v_{\ell,n}}) \rfloor$ is a safe speed of the Krauss-model (Krauss et al, 1997) that is a solution of the Gipps-equation (Gipps, 1981, 1986):

$$v^{(safe)}\tau + X_d(v^{(safe)}) = g_n + X_d(v_{\ell,n}), \qquad (10)$$

$X_d(u)$ is the distance traveled by the vehicle with an initial speed $u$ at a time-independent deceleration $b$ until it comes to a stop; in the model with the discrete time $X_d(u) = b\tau^2\left(\alpha\beta + 0.5\alpha(\alpha - 1)\right)$, $\alpha$ and $\beta$ are the integer and fractional parts of $u/b\tau$, respectively.

The noise component $\xi_n$ in Eq. (2) that simulates random deceleration and acceleration is applied depending on whether the vehicle decelerates or accelerates, or else maintains its speed:

$$\xi_n = \begin{cases} -\xi_b & \text{if } S_{n+1} = -1 \\ \xi_a & \text{if } S_{n+1} = 1 \\ \xi^{(0)} & \text{if } S_{n+1} = 0, \end{cases} \qquad (11)$$

where

$$\xi^{(0)} = a^{(0)}\tau \begin{cases} -1 & \text{if } r < p^{(0)} \\ 1 & \text{if } p^{(0)} \le r < 2p^{(0)} \text{ and } v_n > 0 \\ 0 & \text{otherwise,} \end{cases} \qquad (12)$$

$\xi_a$, $\xi_b$ are random sources for deceleration and acceleration, respectively:

$$\xi_a = a^{(a)}\tau\theta(p_a - r), \qquad (13)$$

$$\xi_b = a^{(b)}\tau\theta(p_b - r), \qquad (14)$$

$p_a$ and $p_b$ are probabilities of random acceleration and deceleration, respectively; $p^{(0)}$, $a^{(0)}$ are constants; $a^{(a)}$ and $a^{(b)}$ are speed functions; $r = \text{rand}(0,1)$, i.e., this is an independent random value uniformly distributed between 0 and 1; $\theta(z) = 0$ at $z < 0$ and $\theta(z) = 1$ at $z \ge 0$; $S_{n+1}$ denotes the state of vehicle motion determined by formula

$$S_{n+1} = \begin{cases} -1 & \text{if } \tilde{v}_{n+1} < v_n \\ 1 & \text{if } \tilde{v}_{n+1} > v_n \\ 0 & \tilde{v}_{n+1} = v_n. \end{cases} \qquad (15)$$

To simulate driver time delays either in vehicle acceleration or in vehicle deceleration under different traffic situations, $a_n$ and $b_n$ in (4), (5) are taken as stochastic functions

$$a_n = a\theta(P_0 - r_1), \qquad (16)$$

$$b_n = a\theta(P_1 - r_1), \qquad (17)$$

$$P_0 = \begin{cases} p_0 & \text{if } S_n \ne 1 \\ 1 & \text{if } S_n = 1, \end{cases} \qquad (18)$$

$$P_1 = \begin{cases} p_1 & \text{if } S_n \ne -1 \\ p_2 & \text{if } S_n = -1, \end{cases} \qquad (19)$$

where $r_1 = \text{rand}(0,1)$; probabilities $p_0(v)$, $p_2(v)$ are given functions of speed, probability $p_1$ is constant; $1 - P_0$ and $1 - P_1$ are probabilities for time delays in vehicle acceleration and deceleration, respectively.

Eqs. (3)-(7) describe the speed adaptation effect in synchronized flow: Within the gap range $g_{s,n} \leq g_n \leq G_n$ ( $g_{s,n}$ is a safe space gap found from the equation $v_n = v_{s,n}$ ), the vehicle tends to adjust its speed to the preceding vehicle. At a given time-independent speed of the preceding vehicle $v_\ell = \text{const}$, this speed adaptation leads to car following with $v = v_\ell$ at a time-independent space gap. There is an infinity number of these gaps associated with the same speed $v = v_\ell = \text{const}$. These gaps lie between the synchronization gap and safe gap, i.e., there is no desired (or optimal) space gap in synchronized flow.

The speed adaptation effect is associated with a driver behavioral assumption of three-phase traffic theory that in hypothetical steady states of synchronized flow (in a steady state all vehicles move at the same time-independent speed and at the same space gap to each other) the driver accepts an infinite number of space gaps at the same speed. These states cover a two-dimensional region in the flow-density plane (dashed region on Fig. 1(c, d)), i.e., there is no fundamental diagram for these steady states. In contrast with the continuum model, in the discrete model the speed and space gap are integer. For this reason, the steady states do not form a continuum in the flow-density plane as they do in the continuum model. The inequalities

$$g \leq G \quad and \quad v \leq \min(v_{\text{free}}, v_s(g, v)) \tag{20}$$

define a 2D-region in the flow-density plane in which the steady states exist (Fig. 1(c, d)).

In a two-lane model used, lane changing occurs between both lanes of the main road independent of whether vehicles are outside or within on- and off-ramp merging regions. A vehicle changes the lane with probability $p_c$, if some necessary(incentive) rules for lane changing from the right lane to the left (passing) lane ( $R \rightarrow L$ ) or from the left to the right lane ( $L \rightarrow R$ ) together with some safety conditions are satisfied. The necessary lane changing rules are

$$R \rightarrow L: \quad v_n^+ \geq v_{\ell,n} + \delta_1 \text{ and } v_n \geq v_{\ell,n}, \tag{21}$$

$$L \rightarrow R: \quad v_n^+ > v_{\ell,n} + \delta_1 \text{ or } v_n^+ > v_n + \delta_1. \tag{22}$$

The safety conditions for lane changing are *either*

$$g_n^+ > \min(v_n \tau, G_n^+), \tag{23}$$

$$g_n^- > \min(v_n^- \tau, G_n^-), \tag{24}$$

$$G_n^+ = G(v_n, v_n^+), \quad G_n^- = G(v_n^-, v_n) \tag{25}$$

*or* (when conditions (23), (24) are not satisfied)

$$x_n^+ - x_n^- - d > g_{\text{target}}^{(\min)}, \tag{26}$$

$$g_{\text{target}}^{(\min)} = \lfloor \lambda v_n^+ + d \rfloor. \tag{27}$$

In addition to (26), the condition that the vehicle passes the midpoint

$$x_n^{(m)} = \lfloor (x_n^+ + x_n^-)/2 \rfloor \tag{28}$$

between two neighboring vehicles in the target lane for time step $n$ should be satisfied, i.e.,

$$x_{n-1} < x_{n-1}^{(m)} \text{ and } x_n \geq x_n^{(m)} \text{ or } x_{n-1} \geq x_{n-1}^{(m)} \text{ and } x_n < x_n^{(m)}. \tag{29}$$

After lane changing the speed $v_n$ is set to

$$\hat{v}_n = \min(v_n^+, v_n + \Delta v^{(1)}); \tag{30}$$

In (30), the speed $v_n$ is related to the lane before lane changing. The vehicle coordinate does not change after lane changing under conditions (23), (24), whereas under condition (26) the vehicle co-ordinate is set to $x_n = x_n^{(m)}$ after lane changing. In (23)-(35), functions $G_n^+$, $G_n^-$ are given by (7); superscripts + and – in variables and functions denote the preceding vehicle and the trailing vehicle in the target lane, respectively; $\Delta v^{(1)}$, $p_c$, $\delta_1$ are constants. In the model, there is an on-ramp bottleneck whose model and parameters have been explained in (Kerner, 2004a).

### 2.3. Some Driver Behavioral Assumptions of the Model

### 2.3.1. In 2D-region of synchronized flow, driver recognizes a change in space gap over time even if the speed difference to the preceding vehicle is negligible

In synchronized flow of a given time-independent speed, a driver accepts a range of the infinite numbers of space gaps to the preceding vehicle (the fundamental hypothesis of three-phase traffic theory). This means that there is no fundamental diagram for steady states of synchronized flow: There are the infinite numbers of hypothetical steady model states of synchronized flow in which vehicles follow each other at the same time-independent speed covering a 2D-region in the flow-density plane (Fig. 1(c)). The 2D-region of steady states is associated with a driver behavioral assumption of the model that in synchronized flow a driver is able to recognize whether the space gap is increasing or deceasing independent of how small the absolute value of the speed difference $\Delta v_n = v_{n,\ell} - v_n$ is. The boundaries of this 2D-region denoted in Fig. 1(c) by F, $S_{low}$, and $S_{upper}$ are respectively associated with free flow, the synchronization space gap $G$ (7), and a safe space gap $g_{safe}$ determined through the safe speed (8) (Fig. 1(d)).

### 2.3.2. Driver accelerates at the downstream jam front with a time delay that is greater than the safe time headway and smaller than the synchronization time headway

At a given steady speed in synchronized flow (this speed is always higher than zero, $v > 0$) a driver behavioral assumption is that $\tau_{del,\,jam}^{(a)}$ is greater than the safe time headway $\tau_{safe} = g_{safe}\,/\,v$ and it is smaller than the synchronization time headway $\tau_G = G\,/\,v$, i.e.,

$$\tau_G > \tau_{del,\,jam}^{(a)} > \tau_{safe}\,. \qquad (31)$$

This condition is equivalent to the hypothesis of three-phase traffic theory of Sect. 2.1: The line *J* divides the 2D-region of steady states of synchronized flow into two classes: the states on and above the line *J* and the states below the line *J*, which are metastable and stable states with respect to wide moving jam formation, respectively (Kerner, 1998).

### 2.3.3. Speed adaptation effect: Driver adapts its speed to the preceding vehicle within the 2D-region of synchronized flow states independent of a space gap

A driver behavioral assumption that results from 2D-region of steady states of synchronized flow (Fig. 1(a)) means that when the vehicle cannot pass the preceding vehicle, within the space gap range

$$G_n > g_n > g_{n,safe} \qquad (32)$$

the vehicle tends to adjust its speed to the preceding vehicle without caring, what the precise space gap is. There is the infinite number of these gaps associated with the same speed $v = v_\ell$, i.e., there is no desired (or optimal) space gap in steady states of synchronized flow.

*2.3.4. Driver searches for the opportunity to accelerate and to pass while moving within the 2D-region of synchronized flow*

In synchronized flow, a driver searches for the opportunity to accelerate and to pass. This driver behavioral assumption is called the over-acceleration effect (Kerner, 2004a). A competition between the speed adaptation (Sect. 2.2.3) and over-acceleration effects simulates traffic breakdown (F → S transition). The over-acceleration is simulated in the model through lane changing to a faster lane (21)-(30) (see Sect. 11.3.3.4 of Kerner, 2009).

*2.3.5. Driver comes on average closer to the preceding vehicle over time while moving in the 2D-region of synchronized flow*

Moving in synchronized flow, a driver comes on average closer to the preceding vehicle over time. This driver behavioral assumption should explain the pinch effect, i.e., the emergence of growing narrow moving jams in synchronized flow. A driver time delay in deceleration simulates this effect through model fluctuations in deceleration $b_n$ (17) that is applied under condition (32) only.

*2.3.6. Over-deceleration effect: Driver decelerates with a delay (reaction time) after the preceding vehicle has begun to decelerate*

This driver behavioral assumption introduced by Herman et al., 1959 and Gazis et al., 1961 is simulated as a collective effect through the use of random fluctuations in vehicle deceleration $\xi_b$ (14), which is applied only if the vehicle should decelerate without model fluctuations. In the model, a competition between the over-deceleration and the speed adaptation effects (Sect. 2.2.3) determines moving jam emergence in synchronized flow.

*2.3.7. Driver accelerates with a time delay after the preceding vehicle has begun to accelerate*

This well-known driver behavioral assumption should describe driver delay in acceleration at the downstream front of synchronized flow or wide moving jam (in the latter case, this driver delay in acceleration is known as a slow-to-start rule) after the preceding vehicle has begun to accelerate. The driver time delay in acceleration is simulated as a collective effect through the use of a random value of vehicle acceleration $a_n$ (16) that is applied under condition (33) and only then if the vehicle did not accelerate at the former time step. The mean time in vehicle acceleration at the downstream jam front is

$$\tau_{\text{del, jam}}^{(a)} = \tau/p_0|_{v_n=0}. \qquad (33)$$

A possible correspondence of the abovementioned driver behavioral assumptions of the model to human factors and psychological insights is out of the paper scope; this can be a very interesting subject of further investigations.

*2.4. Diagram of Congested Patterns at On-Ramp Bottleneck*

A diagram of congested patterns at an on-ramp bottleneck represents types of spatiotemporal congested traffic patterns, which appear spontaneously (Fig. 3(a)) or can be induced (Fig. 3(b)) at the bottleneck[2], in the flow-flow plane whose coordinates are the flow rate upstream of the bottleneck $q_{in}$ and on-ramp inflow rate $q_{on}$. The diagram of congested patterns presented in Fig. 3 is related to model parameters of Table 1 that are usually used in the stochastic model of Sect. 2.2 (Kerner and Klenov, 2009).

---------

[2] We present the boundaries for traffic breakdown both at small $F_S^{(B)}$ (Fig. 3(a)) and great fluctuations $F_{th}^{(B)}$ (Fig. 3(b)), i.e., when free flow is initially at the bottleneck. However, the boundary $S_J^{(B)}$ for wide moving jam emergence shown in Figs. 3(a, b) is associated with small fluctuations in synchronized flow only. In the diagram, there are also boundaries for wide moving jam emergence related to great fluctuations that are not shown for a simplification of the further analysis (see Fig. 18.18 (b, c) in Kerner, 2004a).

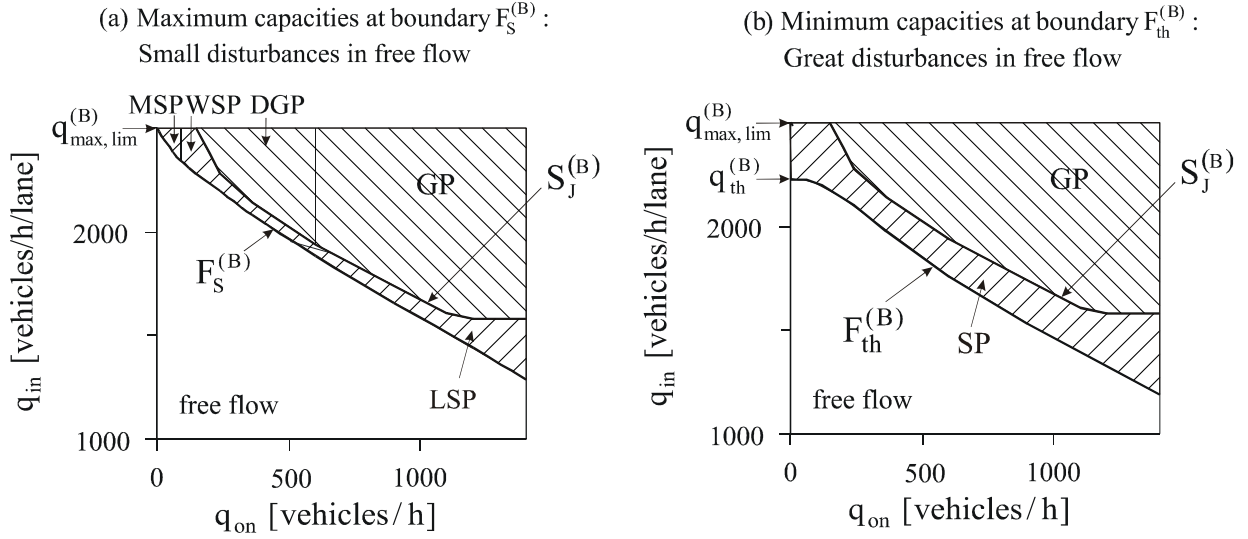

Fig. 3. Diagram of congested patterns at on-ramp bottleneck: (a) For small fluctuations in free flow. (b) For great fluctuations in free flow (we do not show regions of different SPs in this figure). Taken from (Kerner, 2004a).

Table 1: Usual model parameters used in simulations shown in Fig. 3

| Vehicle motion in road lane |
| --- |
| $d = 7.5 \, \text{m}/\delta x$, $\delta x = 0.01 \, \text{m}$, $\delta v = 0.01 \, \text{ms}^{-1}$, $\delta a = 0.01 \, \text{ms}^{-2}$, $v_{\text{free}} = 30 \, \text{ms}^{-1}/\delta v$, $b = 1 \, \text{ms}^{-2}/\delta a$, $a = 0.5 \, \text{ms}^{-2}/\delta a$, $k = 3$, $\phi_0 = 1$, $p_1 = 0.3$, $p_b = 0.1$, $p^{(0)} = 0.005$, $p_2(v_n) = 0.48 + 0.32\Theta(v_n - v_{21})$, $p_0(v_n) = 0.575 + 0.125 \min(1, v_n/v_{01})$, $a^{(b)}(v_n) = 0.2a + 0.8a \max(0, \min(1, (v_{22} - v_n)/\Delta v_{22}))$, $a^{(a)} = 0$, $a^{(0)} = 0.2a$, $v_{01} = 10 \, \text{ms}^{-1}/\delta v$, $v_{22} = 12.5 \, \text{ms}^{-1}/\delta v$, $\Delta v_{22} = 2.778 \, \text{ms}^{-1}/\delta v$, $p_a = 0$, $v_{21} = 15 \, \text{ms}^{-1}/\delta v$. |
| Lane changing parameters |
| $\delta_1 = 1 \, \text{ms}^{-1}/\delta v$, $L_a = 150 \, \text{m}/\delta x$, $p_c = 0.2$, $\lambda = 0.75$, $\Delta v^{(1)} = 2 \, \text{ms}^{-1}/\delta v$. |

There are two main boundaries in the diagram shown in Fig. 3(a):

(i)  At the boundary $F_S^{(B)}$ traffic breakdown occurs spontaneously, i.e., from model fluctuations during a given time of the observation of traffic flow at the bottleneck $T_{ob}$ ($T_{ob} = 40$ min in Fig. 3) at given time-independent values of $q_{in}$ and $q_{on}$.

(ii) At the boundary $S_J^{(B)}$ wide moving jams emerge spontaneously in synchronized flow during a given time of the observation of synchronized flow at the bottleneck $T_{ob}$ ($T_{ob} = 60$ min in Fig. 3).

There are two main types of congested patterns that appear due to traffic breakdown: A synchronized flow pattern (SP) in which congested traffic consists of synchronized flow only and a general pattern (GP) in which congested traffic consists of the two phases, synchronized flow and wide moving jam (Fig. 4). Moreover, there are many different types of SPs (WSP, LSP, MSP) and GPs whose definitions and explanations can be found in (Kerner, 2009).

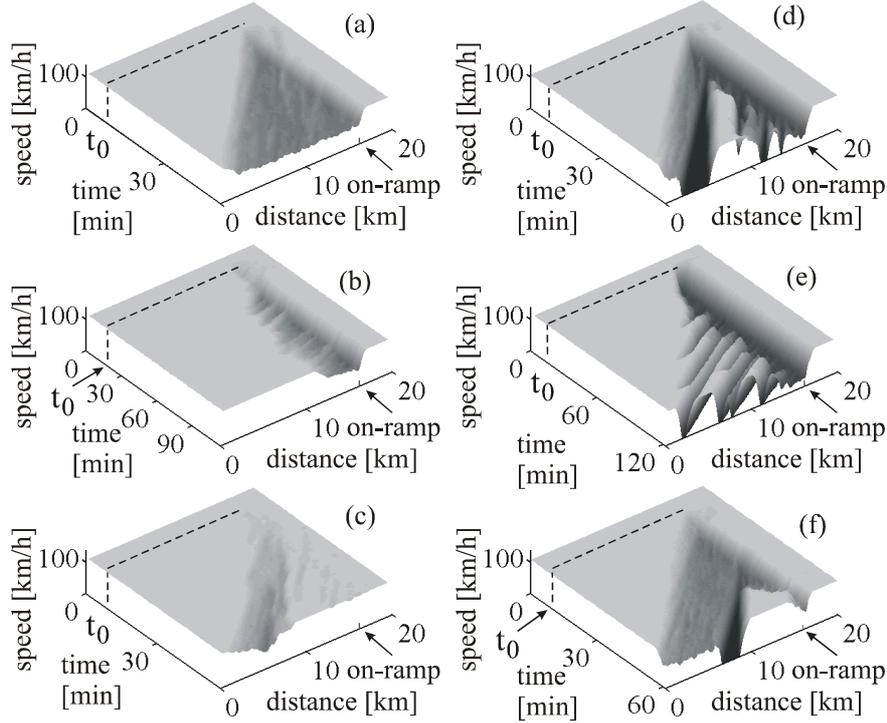

Fig. 4. Congested pattern types related to the diagram shown in Fig. 3(a): (a) Widening SP (WSP). (b) Localized SP (LSP). (c) Moving SP (MSP). (d, e) GPs. (f) Dissolving GP (DGP). Model parameters of Table 1. Taken from (Kerner, 2004a).

There is a threshold boundary for traffic breakdown $F_{th}^{(B)}$ at the diagram (Fig. 3(b)): At the boundary traffic breakdown can still be induced by a great enough time-limited disturbance in free flow at the bottleneck. The region on and between the boundaries $F_{th}^{(B)}$ and $F_S^{(B)}$ traffic breakdown is possible. In three-phase traffic theory, it has been explained why the boundaries $F_{th}^{(B)}$ and $F_S^{(B)}$ are associated with the minimum and maximum highway capacities, respectively (see Sect. 8.3 of Kerner, 2004a).

## 3. Evolution of Congested Patterns under stronger Speed Adaptation within Synchronization Distance

### 3.1. Simulation of Driver Behaviors associated with Speed Adaptation

As mentioned in Introduction, we limit a consideration of the influence of driver behavior associated with driver speed adaptation within 2D-region of synchronized flow states on congested pattern characteristics. This speed adaptation is realized when condition (32) is satisfied. Then the vehicle speed is lower then the safe speed. Let us assume that the speed is higher than the speed of the preceding vehicle

$$v_{s,n} > v_n > v_{\ell,n} > 0,$$ (34)

and the vehicle decelerates at time step n, i.e., the state of vehicle motion in our model $S_n = -1$. For simplification of formulae, we suggest that at next time step $n+1$ speeds $\tilde{v}_{n+1}, v_{n+1} > 0$. Then in accordance with (2)-(19), we find

$$v_{n+1} = \tilde{v}_{n+1} + \xi_n, \tag{35}$$

where

$$\tilde{v}_{n+1} = v_n + \max(-b_n \tau, -(v_n - v_{\ell,n})), \tag{36}$$

$$b_n = a\theta(p_2 - r_1). \tag{37}$$

Formulae (35)-(37) together with (14) describe speed adaptation within 2D-region of synchronized flow under conditions (32). The sense of this speed adaptation is as follows. There are two different driver behavior effects:

1. A driver comes on average closer to the preceding vehicle over time (Sect. 2.3.5).
2. Over-deceleration effect: Driver decelerates with a delay (reaction time) after the preceding vehicle has begun to decelerate (Sect. 2.3.6).

The first driver behavior is simulated in the stochastic model as follows. From (37) we see that with probability $p_2$, deceleration $b_n = a$, i.e., from (36) we get

$$\tilde{v}_{n+1} = v_n + \max(-a\tau, -(v_n - v_{\ell,n})). \tag{38}$$

Due to condition (34), from (38) it follows that the value $\max(-a\tau, -(v_n - v_{\ell,n})) < 0$, i.e.,

$$\tilde{v}_{n+1} < v_n. \tag{39}$$

Therefore, as follows from (15) the state of vehicle motion

$$S_{n+1} = -1. \tag{40}$$

Then in (11) the noise component $\xi_n = -\xi_b$ and formula (35) yields

$$v_{n+1} = v_n + \max(-a\tau, -(v_n - v_{\ell,n})) - \xi_b, \tag{41}$$

i.e.,

$$v_{n+1} < v_n. \tag{42}$$

Thus as follows from (42), with probability $p_2$ at time step $n+1$ the vehicle continues to decelerate trying to approach the speed of the preceding vehicle.

However, with probability $1 - p_2$ from (37) we find that deceleration $b_n = 0$, i.e., instead of (38) from (36) we get

$$\tilde{v}_{n+1} = v_n + \max(0, -(v_n - v_{\ell,n})) = v_n. \tag{43}$$

Then from (15) we find that the state of vehicle motion $S_{n+1} = 0$ and therefore from (11), (35), and (36) instead of (41) we get

$$v_{n+1} = v_n + \xi^{(0)}. \tag{44}$$

This means that without taking into account the noise component $\xi^{(0)}$ (12), with probability $1 - p_2$ the vehicle interrupts its deceleration. This occurs although the vehicle speed is higher than the preceding vehicle. This deceleration interruption describes on average the following: effect: Time headways of some drivers come closer to the boundary of metastable HSFs associated with the safe gap $g_{safe}$ (Figs. 1(b, d)).

As mentioned in Sect. 2.1, the closer the values of space (and, therefore, time) gaps to the safe gap, the greater the probability for moving jam emergence in metastable HSFs at the same $T_{ob}$. This explains the importance of the driver behavioral assumption made in the model about deceleration interruption within the 2D-region of synchronized flow described by the value of probability $1 - p_2$ for the deceleration interruption.

The second driver behavior is simulated in the stochastic model as follows. There is probability $p_b$ for random fluctuations in vehicle deceleration $\xi_b$ (14). As follows from Eq. (41), the value $\xi_b$ influences considerably on the speed $v_{n+1}$, specifically on the vehicle deceleration. To understand the impact of probability $p_b$ on moving jam emergence in metastable HSFs, we note that from (14) it follows that with probability $p_b$ the value of speed fluctuation $\xi_b = a^{(b)}\tau$ and from (41) we get

$$v_{n+1} = v_n + \max(-a\tau, -(v_n - v_{\ell,n})) - a^{(b)}\tau, \qquad (45)$$

i.e., due to the speed fluctuation, the speed decreases additionally in comparison with "deterministic" speed deceleration within 2D-region of synchronized flow described by Eq. (38). This random addition speed decrease describes on average the well-known over-deceleration effect of the GM model class mentioned in Sect. 2.3.6. Otherwise, with probability $1 - p_b$ from (14) we find that $\xi_b = 0$, i.e., there is no additional random deceleration and the speed adaptation is associated with only the "deterministic" deceleration within 2D-region of HSFs; for this case, from (41) we get

$$v_{n+1} = \tilde{v}_{n+1}, \qquad (46)$$

where the "deterministic" speed component $\tilde{v}_{n+1}$ is given by Eq. (38). Thus the greater the probability $p_b$ of random deceleration is, the stronger the over-deceleration within 2D-region of synchronized flow.

Thus we can conclude that

1. The driver behavior to adapt its speed to the speed of the preceding vehicle moving within 2D-region of synchronized flow is described in the model by probability $p_2$ for continuous deceleration to the speed of the preceding vehicle: The greater the probability $p_2$, the stronger on average the speed adaptation of the vehicle to the speed of the preceding vehicle.

2. The driver over-deceleration is described in the model by probability $p_b$ of random deceleration: The smaller the probability $p_b$, the weaker the driver over-deceleration , i.e., the smaller the mean driver delay in deceleration.

Below we consider the impact of these driver behaviors on congested traffic patterns at an on-ramp bottleneck.

### 3.2. Impact of Driver Behavior on Diagram of Congested Patterns at Bottleneck

Firstly, we consider how the diagram of congested patterns changes, when we increase $p_2$ and decrease $p_b$ in comparison with their values used in Fig. 3 (a) given in Table 1 (see changed model parameters in caption to Fig. 5).

The changes in driver behavior lead to a considerable shift of the boundary $S_J^{(B)}$ to the right in the diagram. At the boundary $S_J^{(B)}$ wide moving jams emerge spontaneously in synchronized flow during the observation time $T_{ob} = 60$ min (Fig. 5). Due to this boundary shift, at the flow rate upstream of the bottleneck $q_{in}$ and on-ramp inflow rate $q_{on}$ associated with points B and C in Fig. 5, at which GPs emerge at the bottleneck in the diagram shown in Fig. 3, SPs occur under changed driver characteristics (Fig. 6).

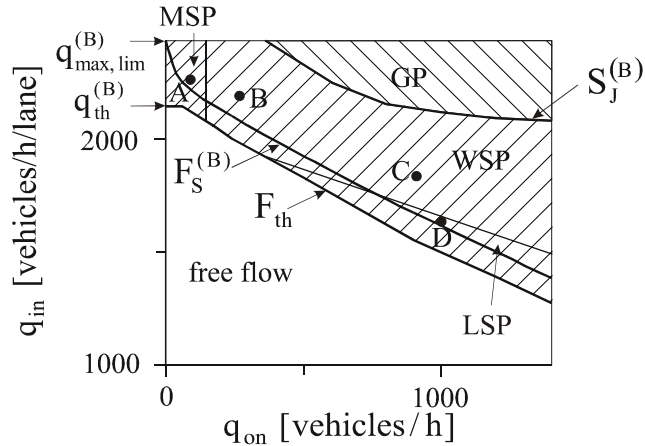

Fig. 5. Congested pattern diagram at $p_2(v_n) = 0.75 + 0.25\Theta(v_n - v_{21})$ and $p_b = 0.05$. Other model parameters are the same as those in Table 1.

This occurrence of SPs at the bottleneck (Fig. 6 (c, d)) at relatively great flow rates $q_{in}$ and $q_{on}$ can be explained as follows. When the speed adaptation of the vehicle to the speed of the preceding vehicle becomes stronger ($p_2$ is greater), the percentage of vehicles moving in synchronized flow at space gaps that are close to a safe gap decreases. For this reason, synchronized flow becomes more stable with respect to moving jam formation, i.e., probability of wide moving jam emergence in synchronized flow decreases. In addition, due to a decrease in probability $p_b$ of random deceleration, the over-deceleration effect in synchronized flow, which causes the nucleus occurrence for wide moving jam emergence, becomes weaker. Thus whether an GP or SP occurs at chosen flow rates $q_{in}$ and $q_{on}$ depends considerably on driver behavioral characteristics under consideration.

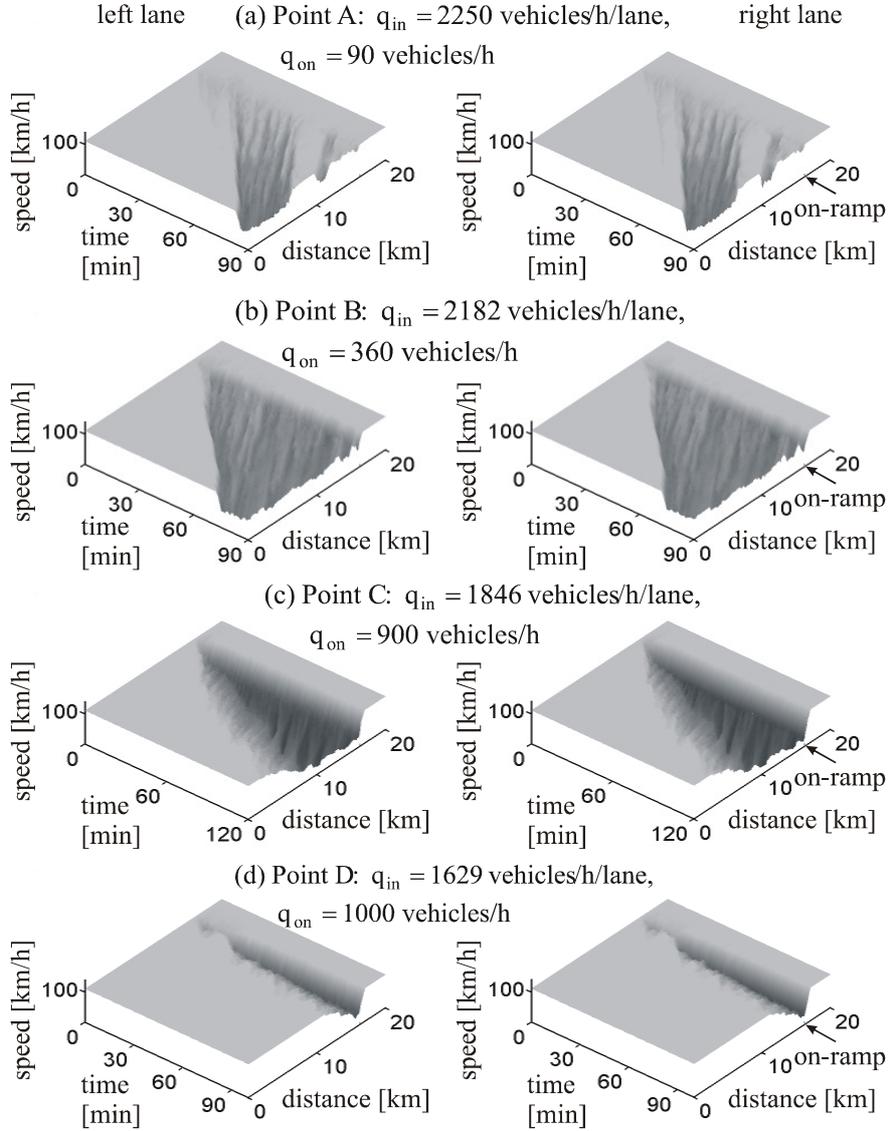

Fig. 6. Congested patterns associated with points A-D in the diagram shown in Fig. 5.

### 3.3. A Variety of Different Congested Patterns randomly Occurring at the same Flow Rates at Bottleneck

As mentioned, at the boundary $S_J^{(B)}$ in diagram shown in Fig. 5, a wide moving jam emerges spontaneously in synchronized flow with probability 1 during $T_{ob} = 60$ min. This probability has been found from simulations of 40 different realizations in which model parameters and the flow rates are the same, however, *initial conditions* for traffic variables associated with random numbers $r = \text{rand}(0,1)$ in the model of Sect. 2.2 are different.

Because the emergence of a wide moving jam in synchronized flow is a random event, we can expect that with probability that is smaller than 1 a wide moving jam(s) can emerge in some of the 40 realizations even at the flow rates $q_{in}$ and $q_{on}$ associated with points in the diagram *left* to the boundary $S_J^{(B)}$. This means with some probability rather than an SP a general pattern should occur at least in some diagram region between the boundaries $S_J^{(B)}$ and $F_S^{(B)}$ (Kerner, 2004a). This effect is easily realized at stronger speed adaptation within 2D-region of synchronized flow under consideration. This is explained by a very broad region of the flow rates $q_{in}$ and $q_{on}$ at which SPs can occur and exist in the case of the stronger speed adaptation (Figs. 5 and 6).

Results of this statistical analysis of a variety different congested patterns occurring at the *same* flow rates, vehicle and bottleneck characteristics are presented in Figs. 7-9. We find that in different realizations either an WSP (Fig. 8 (a)) or an GP (Fig. 8 (b)) occur. This emphasizes the stochastic nature of wide moving jam emergence in synchronized flow in three-phase traffic theory (Kerner, 2004a):

- At the same flow rates at a bottleneck, the same vehicle and driver characteristics in traffic flow, and the same bottleneck characteristics qualitatively different either SPs or GPs can occur at the bottleneck. In this case, the type of the pattern and its parameters depends on *random* disturbances occurring in traffic flow. This confirms one of the important general results of three-phase traffic theory (Kerner, 2004a).

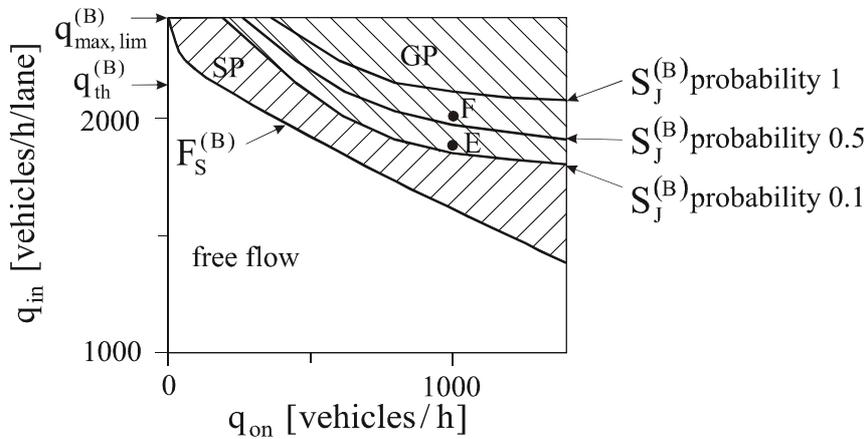

Fig. 7. Congested pattern diagram shown in Fig. 5 with two addition boundaries $S_J^{(B)}$ associated with probability of GP formation 0.5 and 0.1. The boundary $S_J^{(B)}$ related to probability 1 is the same as the boundary $S_J^{(B)}$ in Fig. 5.

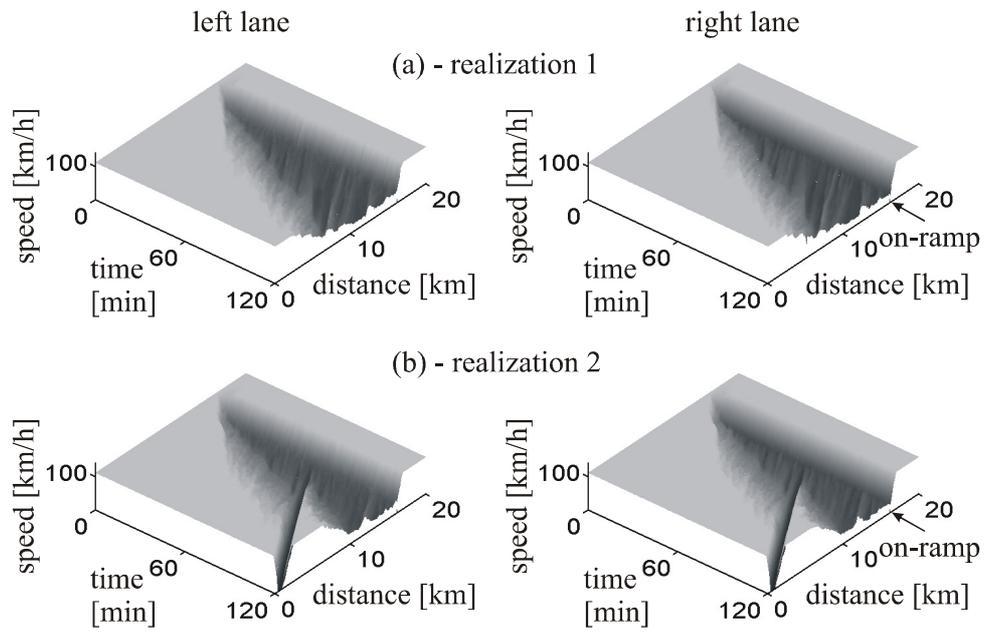

Fig. 8. Two different realizations of congested patterns associated with the point E in the diagram shown in Fig. 7. $q_{in}$ = 1846 vehicles/h/lane, $q_{on}$ = 1000 vehicles/h.

This general conclusion of three-phase traffic theory can also be seen in Fig. 9 in which three different congested patterns occur in three realizations simulated at the same flow rates, the same vehicle and driver characteristics in traffic flow, and the same bottleneck characteristics.

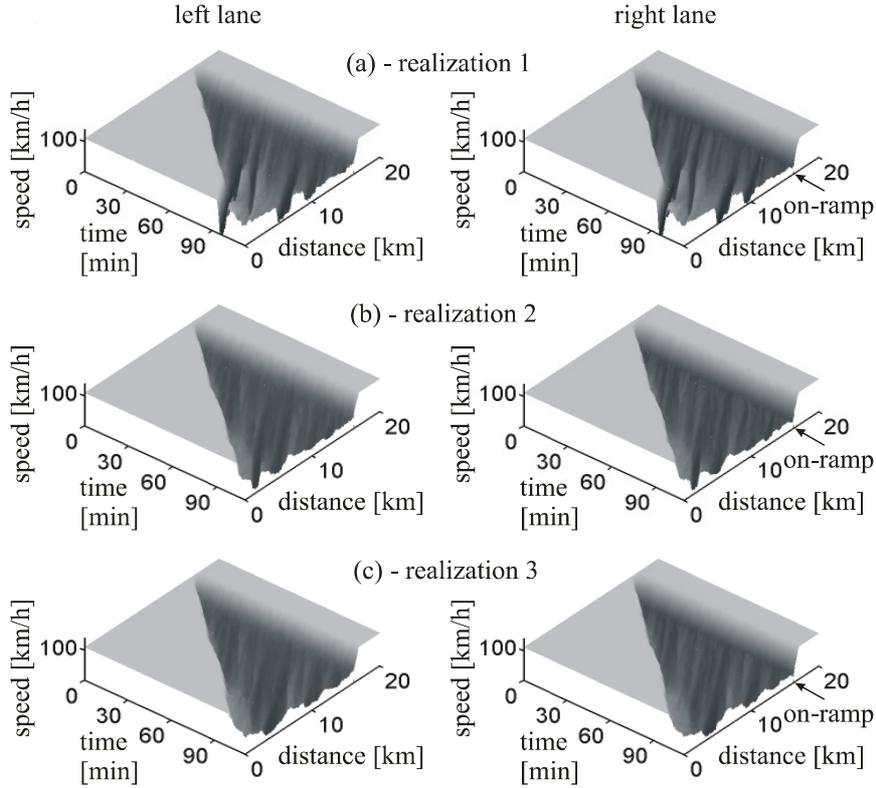

Fig. 9. Three different realizations of congested patterns associated with the same point F in the diagram shown in Fig. 7. $q_{in}$ = 2000 vehicles/h/lane, $q_{on}$ = 1000 vehicles/h.

### 3.4. Widening Synchronized Flow Patterns (WSP) at Great Bottleneck Strengths resulting from the Fundamental Hypothesis of Three-Phase Traffic Theory

What does happen with congested traffic patterns at the bottleneck, if drivers increase the speed adaptation effect within 2D-region of synchronized flow under condition (32) further? The answer on this question is shown in Fig. 10. Here we choose the "full" speed adaptation within 2D-region of synchronized flow associated with $p_2(v_n) = 1$ and decrease probability of random deceleration to $p_b = 0.005$. Then we find that no wide moving jams emerge in synchronized flow at all flow rates $q_{in}$ and $q_{on}$ shown in the diagram in Fig. 10 (a).

In other words, only different SPs occur at the bottleneck at such "full" speed adaptation within 2D-region of synchronized flow (Fig. 10 (b-d)). This result follows from the fundamental hypothesis of three-phase traffic theory: There are the infinite numbers of HSFs (Fig. 2) in which wide moving jams should not necessarily emerge. These HSFs can be even at great vehicle densities and associated low speeds. This is because in three-phase traffic theory *none* of the HSFs within the 2D-region of HSFs (Fig. 2) are unstable with respect to infinitesimal disturbances in synchronized flow (Kerner, 1998).

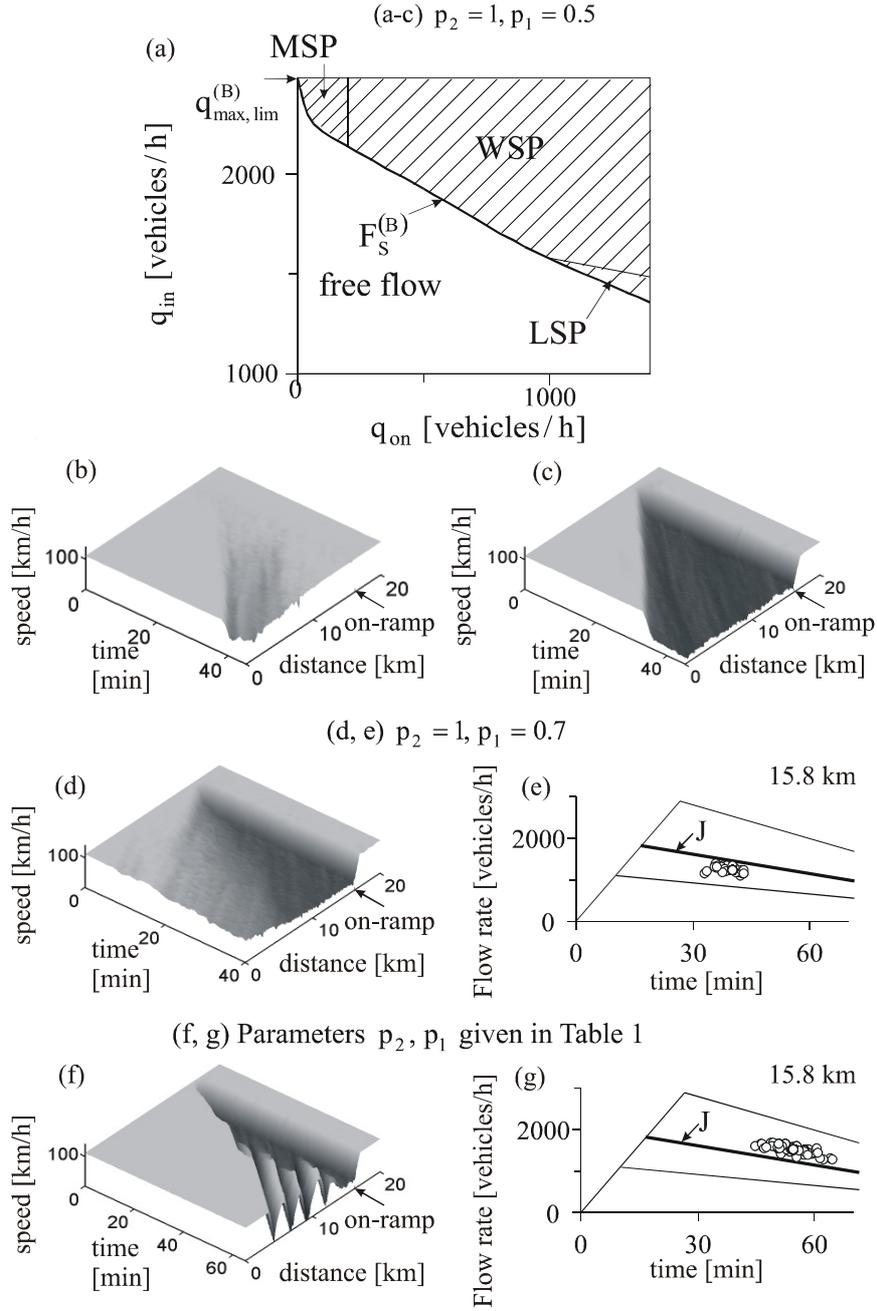

Fig. 10. Features of congested patterns at great speed adaptation: (a-c) Congested pattern diagram (a) at $p_2(v_n) = 1$ and $p_1 = 0.5$ and related MSP (b) and WSP (c). (d, e) WSP (d) and data within synchronized flow (e) at $p_2(v_n) = 1$ and $p_1 = 0.7$. Other model parameters in (a-e) are the same as those in Table 1. (f, g) GP (f) and data within synchronized flow (g) at model parameters of Table 1. In (b-d), (f) speed in space and time in the right lane. In (b-g), $(q_{on}, q_{in}) = (30, 2118)$ (b), $(1400, 2105)$ (c), $(1500, 1846)$ (d-g) ( vehicles/h, vehicles/h/lane). Average (30 min interval) time headway in synchronized flow of WSP (d, e) is 2.04 s, i.e., is longer than $\tau_{del, jam}^{(a)} = 1.74$ s, while for pinch region of GP (f, g) the average time headway is about 1.39 s, i.e., it is shorter than $\tau_{del, jam}^{(a)}$.

The hypothesis about great density HSFs incorporated in the traffic flow model explains results of simulations presented in Fig. 10 in which widening synchronized flow patterns (WSP) of a great density and low speed (e.g., Fig. 10(d)) appear upstream of a bottleneck. Such great density and low speed WSPs exist at great bottleneck strengths, when speed adaptation of drivers is great enough (Fig. 10(a)): In this case, dynamic synchronized flow within an WSP is associated mostly with stable synchronized flow states that are below the line $J$ (Fig. 2). Thus in three-phase traffic theory

- WSPs can occur even at heavy bottlenecks. In particular, this effect can be found when most drivers move in synchronized flow at time headways that are longer than $\tau_{\mathrm{del,\,jam}}^{(a)}$ (Fig. 2). In other words, WSPs at heavy bottlenecks can be found in real measured traffic data [3].

The importance of 2D-region of synchronized flow for the explaining of real measured data (Kerner, 2004a, 2009) is one of the basic results for the introduction by the author of "driver alike adaptive cruise control" (DA-ACC) (Kerner, 2004b), i.e., the ACC in the framework of three-phase traffic theory briefly discussed below.

## 4. Driver Alike Adaptive Cruise Control (DA-ACC) – ACC in the Framework of Three-Phase Traffic Theory

### 4.1. Basic Operation Modes of ACC

Before we consider DA-ACC, we explain the main operation mode of a usual ACC widely used in vehicles (Fig. 11) (see references in Chap. 32 of Winner et al., 2009). The ACC-vehicle measures the space gap $g$ and speed difference between the preceding vehicle and ACC-vehicle $\Delta v = v_\ell - v$, where $v$ is the ACC-vehicle speed, $v_\ell$ is the speed of the preceding vehicle. Based on the current values of $g$, $v$, and $\Delta v$, the ACC vehicle calculates the current time headway $\tau_{\mathrm{headway}}$ between the ACC-vehicle and the preceding vehicle. For simplicity, we consider here only a case in which absolute values of $\Delta v$, the difference $\Delta \tau = \tau_{\mathrm{headway}} - \tau_{\mathrm{ACC}}$ (where $\tau_{\mathrm{ACC}}$ is a desired time headway chosen by a driver), and acceleration (deceleration) of the preceding vehicle are not very great.

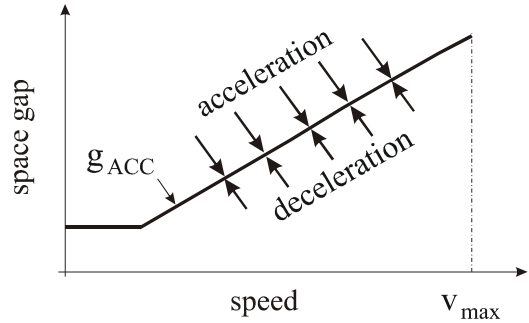

Fig. 11. Main operation mode of a usual ACC (see references in Chap. 32 of Winner et al., 2009)

Then the acceleration (deceleration) of the ACC-vehicle is given by a well-known formula:

$$a_{\mathrm{ACC}} = K_1(g(t) - g_{\mathrm{ACC}}) + K_2 \Delta v(t), \qquad (47)$$

---

[3] Up to now empirical WSP of a great density at heavy bottlenecks are not found in measured traffic data: As proven in Sect. 10.3.10 of (Kerner, 2009), an empirical proof of homogeneous congestion made in (Schönhof and Helbing, 2007, 2009) is invalid.

where $g_{ACC}$ is a desired space gap (Fig. 11) related to the desired time headway $\tau_{ACC}$. $K_1$, $K_2$ are dynamic coefficients of ACC ($K_1$, $K_2 > 0$). The ACC should maintain $\Delta v$ close to zero and the space gap g close to $g_{ACC}$, i.e., the time headway should be close to $\tau_{ACC}$. At higher speeds, $g_{ACC} = v\tau_{ACC}$ is an increasing speed function (Fig. 11). When $\Delta v = 0$ and $g > g_{ACC}$, as follows from (47) the ACC-vehicle accelerates; otherwise, i.e., at $g < g_{ACC}$, the ACC-vehicle decelerates (Fig. 11).

### 4.2. Basic Operation Mode of DA-ACC

In accordance with one of the basic driver behavioral assumptions of three-phase traffic theory, when the space gap to the preceding vehicle is within a 2D-region in the space-gap-speed plane (dashed region in Fig. 1 (b)), a driver adapts its speed to the speed of the preceding vehicle without caring, what the precise space gap is (Kerner, 1998, 2004a). For this reason, we call an ACC with this fundamental feature of three-phase traffic theory as a *driver alike ACC* (DA-ACC). Thus the basic operation mode of an DA-ACC vehicle is given by formula

$$a_{DA-ACC} = K_{\Delta v}\Delta v(t) \quad \text{at} \quad G \geq g \geq g_{safe} \tag{48}$$

associated with three-phase traffic theory: Within 2D-region in the space-gap-speed plane (dashed region in Fig. 12) acceleration (deceleration) of the DA-ACC vehicle does *not* depend on the space gap, i.e., on the time headway to the preceding vehicle at all. In other words, in contrast with the basic formula of an ACC-vehicle (47), the DA-ACC mode (48) does not maintain a desired time headway chosen by the driver (Kerner, 2004b).

As in the Kerner-Klenov model, in (48) $G$ and $g_{safe}$ are the synchronization and safe space gaps, respectively; $K_{\Delta v}$ is a dynamic coefficient that is greater than zero. At $g > G$ the DA-ACC-vehicle accelerates, whereas at $g < g_{safe}$ the DA-ACC-vehicle decelerates. In other words, outside of the 2D-region in the space-gap-speed plane formula (48) is not applied. A possible variety of operation DA-ACC modes labeled in Fig. 12 as "acceleration" and "deceleration" outside of the 2D-region are described in (Kerner, 2007).

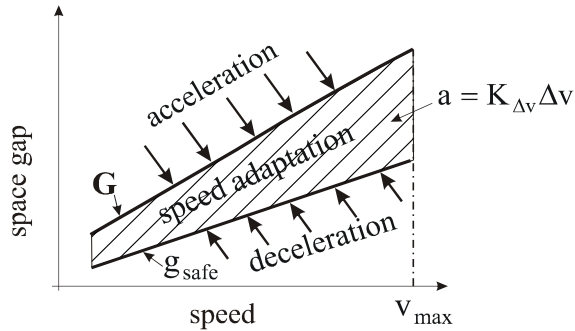

Fig. 12. Basic operation mode of DA-ACC (Kerner, 2004b, 2007, 2009): A part of 2D-region of space gaps between a vehicle moving in accordance with the DA-ACC mode (48) and the preceding vehicles in the space-gap-speed plane (dashed region).

### 4.3. Prevention of Traffic Congestion at Bottleneck through the Use of DA-ACC

Each developer of ACC knows that if dynamic coefficients $K_1$ and $K_2$ of ACC in (47) are chosen great enough, then any platoon of the ACC-vehicles is stable and no moving jam emerge in traffic flow consisting only of the ACC-vehicles. Indeed, when $K_1$ and $K_2$ in (47) are great enough, ACC-vehicles react very quickly on changes in values $(g(t) - g_{ACC})$ and $\Delta v(t)$, therefore, the ACC-vehicles damp speed disturbances within the platoon.

However, such quick damping of speed disturbances at great values of $K_1$, $K_2$ in (47) is very uncomfortable for drivers. This is because the very quick damping of speed disturbances causes considerable ACC-vehicle acceleration (deceleration) even at relatively small values $(g(t) - g_{ACC})$ and $\Delta v(t)$. For this reason, for comfortable driving developers of real ACC have to choose relative small dynamic coefficients $K_1$, $K_2$ of ACC, which cannot prevent moving jam emerge in traffic flow (see for example Sect. 23.6.3 of Kerner, 2004a). The mentioned problem is one the problems of ACC-vehicles leading to a conflict between the dynamic and comfortable ACC behavior.

An DA-ACC-vehicle *removes* this conflict and it *enhances* traffic flow considerably. This is because an DA-ACC is related to a "hypothetical driver" that while approaching a slower moving preceding vehicle is able to perform the "perfect" speed adaptation to the speed of the preceding vehicle under condition (32); in addition, this "hypothetical driver" does not exhibit long time delays that are usual for real drivers.

The effect of DA-ACC-vehicles on traffic flow is shown in Fig. 13. In these simulations, there are DA-ACC-vehicles randomly distributed between vehicles without any ACC. The vehicles without ACC move in accordance with the stochastic model of Sect. 2.2 with model parameters of Table 1.

In the simulation scenario (Fig. 13 (a)), an GP occurs spontaneously at the on-ramp bottleneck when there are *no* DA-ACC-vehicles in traffic flow. When DA-ACC vehicles appear, a strong reduction in traffic congestion is observed (Fig. 13 (b-d)). The greater the percentage of the DA-ACC vehicles is, the less the traffic congestion at the bottleneck (Fig. 13 (b, c)). There is a critical percentage of the DA-ACC vehicles, which is about 28%: when the percentage of the DA-ACC vehicles is greater than the critical one, no traffic breakdown and therefore no traffic congestion occurs at the bottleneck (Fig. 13 (d)).

It must be stressed that the effect of traffic congestion avoidance through the use of DA-ACC is reached at a *small dynamic coefficient* in the basic DA-ACC mode (48) $K_{\Delta v} = 0.14 \text{ s}^{-1}$, which is associated with a very comfortable driving. For a comparison, to prevent moving jam emergence with a usual ACC (47), in simulations presented in Fig. 23.18 of the book (Kerner, 2004a) the associated dynamic coefficient in (47) should be $K_2 = 0.55 \text{ s}^{-1}$ related to a very uncomfortable driving.

The effect of DA-ACC vehicles on traffic flow (Fig. 13 (d)) is explained as follows. The DA-ACC vehicles do not exhibit relatively long time delays in deceleration and acceleration that are usual for drivers. These driver time delays are responsible both for traffic breakdown in free flow and the resulting moving jam emergence in synchronized flow. Therefore in traffic flow without DA-ACC vehicles the GP occurs (Fig. 13 (a)), whereas in traffic flow in which the DA-ACC vehicles dominate free flow remains (Fig. 13 (d)).

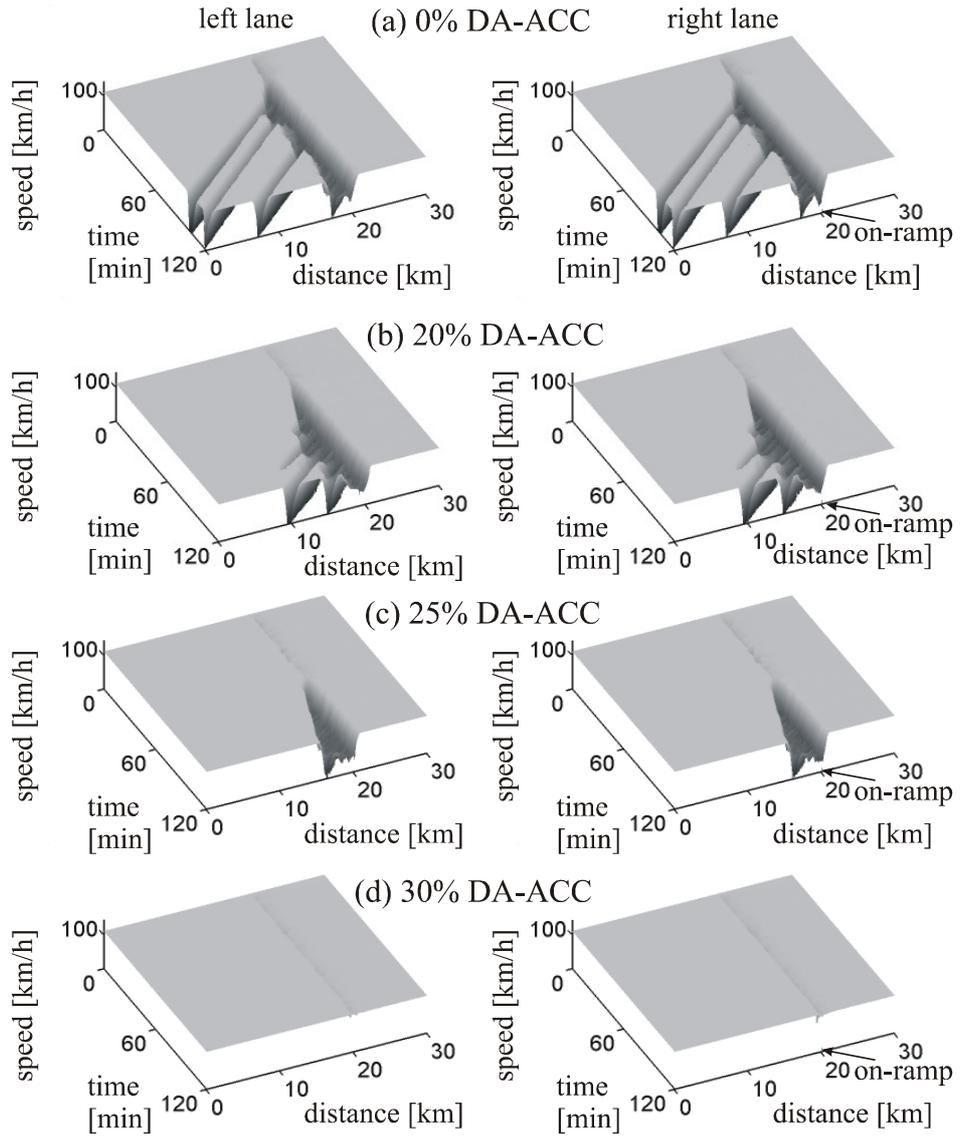

Fig. 13. Effect of DA-ACC on traffic flow: (a) An GP is emerging at an on-ramp bottleneck when all vehicles are usual ones without DA-ACC, i.e., there are no DA-ACC vehicles in traffic flow. (b-d) Traffic patterns at the bottleneck at the same flow rates, however, under different percentages of DA-ACC vehicles that are randomly distributed between usual vehicles without DA-ACC-vehicles. Speed over space and time in the left (left) and right lanes (right). Parameters of the 2D-region for synchronized flow associated with DA-ACC vehicles are  $k = 1.5$ , $\phi_0 = 2$ at $v < v_\ell$ and $\phi_0 = 1$ at $v \geq v_\ell$ .

**5. Discussion**

*5.1. Conclusions*

   Simulations with a stochastic microscopic three-phase traffic flow model allow us to make the following conclusions:

1. Drivers behaviors associated with speed adaptation to the speed of the preceding vehicle within 2D-region of synchronized flow influence traffic congestion at a highway bottleneck considerably. When the speed adaptation is strong enough, no wide moving jams occur in synchronized flow even at great flow rates upstream of the bottleneck. In this case, WSPs remain at the bottleneck at great bottleneck strengths.

2. The greatest speed adaptation to the speed of the preceding vehicle within 2D-region of synchronized flow can be realized due to the use of an ACC in the framework of three-phase traffic theory called as driver alike ACC (DA-ACC). The simulations of DA-ACC behavior explain the following advantages of the DA-ACC (48) based on three-phase traffic theory in comparison with the usual ACC (47):

   • The removing of a conflict between the dynamic and comfortable ACC behavior. In particular, a much comfortable vehicle motion in congested traffic is possible.

   • DA-ACC-vehicles can prevent traffic breakdown at a bottleneck and wide moving jam emergence in synchronized flow.

   • As well-known, under congestion fuel consumption increases; therefore, the DA-ACC that decreases speed disturbances in traffic flow and prevents congestion can lead to reduction in fuel consumption and $CO_2$ emissions.

*5.2. Criticism of "Criticism of Three-Phase Traffic Theory"*

   Three-phase traffic theory is an alternative theory to well-known and widely accepted earlier traffic flow theories and models reviewed in (May, 1090; Leutzbach, 1998; Gartner et al, 2001; Chowdhury, et al, 2000; Helbing, 2001; Nagatani, 2002; Nagel, et al, 2003; Mahnke, et al, 2005; Rakha, et al, 2009). Recently, this theory has been criticized in (Schönhof and Helbing, 2007, 2009). Author's responses to these criticisms have already been done in Sect. 10.3 of (Kerner, 2009). Nevertheless, for a deeper understanding of the article results, we make more detailed explanations of some of these critical responses on (Schönhof and Helbing, 2007, 2009) below.

*5.2.1. Traffic breakdown: F → S transition*

   The main criticism made in three-phase traffic theory (Kerner, 2004a, 2009) on the earlier traffic flow theories including traffic flow models of the group of Helbing (Helbing, 2001; Schönhof and Helbing, 2007, 2009), which explain traffic breakdown by free flow instability (Gazis, 1961; May, 1090; Leutzbach, 1998; Gartner et al, 2001; Chowdhury, et al, 2000; Nagatani, 2002; Nagel, et al, 2003; Mahnke, et al, 2005; Rakha, et al, 2009), is that this free flow instability is responsible for wide moving jam formation in free flow (F → J transition). In contrast, traffic breakdown observed in real measured data is governed by a first-order F → S transition (Kerner, 2004a, 2009).

   In (Schönhof and Helbing, 2007, 2009), an empirical observation of a "boomerang effect", i.e., "growing perturbations on a homogeneous freeway section without on- and off-ramps" (see caption to Figure 1 of Schönhof and Helbing, 2007) that should lead to wide moving jam emergence in free flow has been stated. It must be stressed that the term *boomerang effect* has a sense *only* when traffic breakdown occurs *without influence of a bottleneck*. Otherwise, if a bottleneck is the reason for traffic breakdown and the resulting upstream congestion propagation, then this well-known and usual way of the congestion occurrence has nothing to do with the *boomerang* effect. This "boomerang effect" should prove that a spontaneous F → J transition might govern traffic breakdown in free flow. This should be one of the criticisms of three-phase traffic theory in which rather the spontaneous F → J transition a sequence of F → S → J transitions governs the spontaneous wide moving jam emergence (Kerner, 1998).

However, as proven in (Kerner and Klenov, 2009) based on the *same* measured traffic data as that used in (Schönhof and Helbing, 2007, 2009) in this data rather than the "boomerang effect, an F → S transition occurs at an off-ramp bottleneck resulting in moving SP (MSP) formation; later, due to an S → J transition within the MSP it transforms into a wide moving jam. Thus in the empirical data wide moving jams emerge due to sequence of F → S → J transitions, i.e., the study of (Schönhof and Helbing, 2007, 2009) is *invalid*. A more detailed consideration of the "boomerang effect" in free flow and the associated criticism of these empirical studies can be found in Sect. 10.3.7 of the book (Kerner, 2009).

### 5.2.2. Proof of criticism of HCT and OCT model solutions as well as diagram of congested patterns of Helbing, Treiber et al

In some traffic flow models in the framework of the fundamental diagram hypothesis, the density region at the fundamental diagram, within which traffic flow is unstable (dashed curve in Fig. 14 (a)), is limited at greater densities: At the densities $\rho \geq \rho_{cr}^{(HCT)}$ states on the fundamental diagram are stable with respect to small disturbances. These stable homogeneous in space and time model solutions of a great density, which can appear upstream of a bottleneck in simulations of these models, have been called as HCT in (Helbing, et al, 1999; Treiber, et al, 2000; Helbing, 2001).

In (Kerner, 2004a), the application of HCT model solutions for explanations of real measured data made in (Helbing, 2001; Treiber, et al, 2000) has been criticized. Below based on numerical simulations of features of HCT model solutions, we prove this criticism and show that

- the HCT model solutions, which are the *basic* solutions for the theoretical diagrams of congested patterns derived by Helbing, Treiber et al. (Helbing, et al, 1999; Treiber, et al, 2000, 2010; Helbing, 2001; Schönhof and Helbing, 2007, 2009; Helbing et al., 2009), have *no* sense for real traffic.

To prove that the HCT model solutions have no sense for real traffic flow, we should note that in all known empirical observations, the downstream front of a wide moving jam propagates through any dense congested traffic and bottlenecks while maintaining the mean velocity of the front $v_g$. In other words, this empirical feature [J] of wide moving jams is *independent* of the state of traffic flow downstream of the jam.

However, traffic flow models with the HCT model solutions *cannot* satisfy this very important characteristic empirical feature of traffic. To illustrate this, we consider features of wide moving jam propagation in these models with an example of the fundamental diagram shown in Fig. 14 (a). We find that the characteristic jam feature [J] in these models is satisfied when free flow is downstream of the jam (wide moving jam labeled by "jam A" in Fig. 14 (c, d)). In contrast, the jam feature [J] does *not* remain when an HCT model solution is downstream of a wide moving jam (the jam labeled by "jam B"). This is because rather than the mean front $J$, the downstream front of the "B" propagates with a negative velocity $v_g^{(HCT)}$ that is associated with a line $J^{(HCT)}$ in the flow-density plane between the state within the jam with the jam density $\rho_{max}$ and a point at the fundamental diagram for an HCT solution (with a density $\rho^{(HCT)}$ in Fig. 14 (b)). As a result, the absolute value of the downstream front velocity of the "jam A" $\left| v_g \right|$ is considerably greater than the one for the "jam B" $\left| v_g^{(HCT)} \right|$. Moreover, the greater the density $\rho^{(HCT)}$ of the HCT, the smaller $\left| v_g^{(HCT)} \right|$. As abovementioned, this is inconsistent with measured traffic data of wide moving jam propagation. This proves that HCT model solutions of Ref. (Helbing, 2001; Helbing, et al, 1999; Treiber, et al, 2000, 2010; Schönhof and Helbing, 2007, 2009; Helbing et al., 2009) are not consistent with the empirical feature [J] of wide moving jam propagation through any dense congested traffic while maintaining the mean velocity of the jam front. For this reason, the features of these HCT model solutions have no sense for real traffic flow.

Due to the existence of HCT model solutions, these traffic flow models exhibit also model solutions called oscillating congested traffic (OCT) (Helbing, et al, 1999; Helbing, 2001; Treiber, et al, 2000, 2010; Schönhof and Helbing, 2007, 2009; Helbing et al., 2009). An OCT model solution appears in these models due to the existence of the critical density $\rho_{cr}^{(HCT)}$ for an instability of HCT: When the density in HCT decreases and this HCT density approaches the critical density $\rho_{cr}^{(HCT)}$ (Fig. 14 (a)), then due to HCT instability, OCT model solution occurs. Thus we can make the conclusions:

1.  OCT model solutions of (Helbing, et al, 1999; Helbing, 2001; Treiber, et al, 2000, 2010; Schönhof and Helbing, 2007, 2009; Helbing et al., 2009) appear as a result of the instability of HCT model solutions in traffic flow models with the fundamental diagram shown in Fig. 14(a).
2.  As proven above, HCT model solutions of these models have no sense for real traffic. Therefore, the OCT model solutions resulting from HCT as well as theoretical congested patterns consisting of HCT and OCT combinations have also no relation to real traffic.
3.  HCT and OCT model solutions are important elements of diagrams of congested patterns at bottlenecks of (Helbing, et al, 1999; Helbing, 2001; Treiber, et al, 2000, 2010; Schönhof and Helbing, 2007, 2009; Helbing et al., 2009). This is one of the reasons why these theoretical diagrams have no sense and, therefore, no application to real traffic.

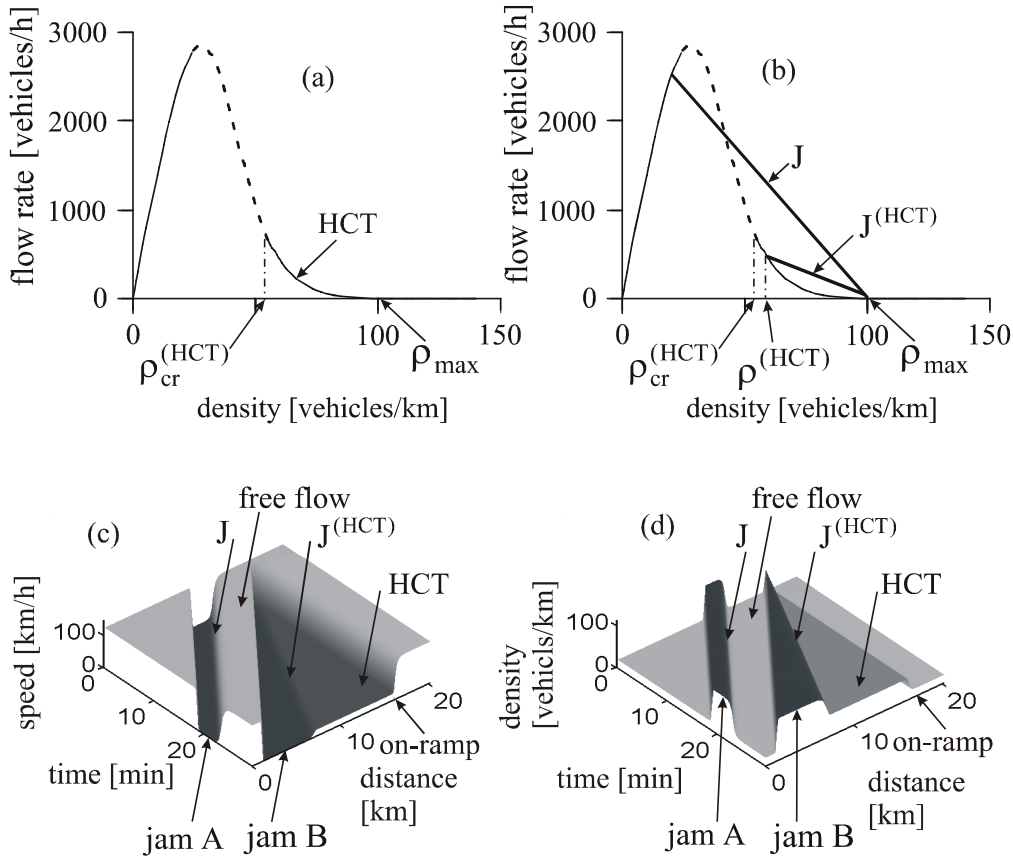

Fig. 14. Simulations of HCT model solutions and wide moving jam propagation: (a) Fundamental diagram with HCT  (b) Fundamental diagram of (a)  together with the lines $J$ and $J^{(HCT)}$  that represent  in the flow--density plane the downstream fronts of two wide moving jams denoted by "jam A" and "jam B", respectively in (c, d). (c, d) Propagation of  two wide moving jams when downstream of the jams either free flow ("jam A") or HCT ("jam B") occur, respectively; speed (b) and density (c) in space and time. Payne-like model. Taken from (Kerner, 2009).

### 5.2.3. Criticism of explanations of empirical congested patterns through spatiotemporal combinations of HCT and OCT model solutions

We should note that in simulations of non-linear traffic flow models of (Treiber, et al, 2000, 2010; Schönhof and Helbing, 2007, 2009), a diverse variety of spatiotemporal combinations of the HCT and OCT model solutions can be found. Through an appropriate choice of boundary and initial conditions used in simulations, some of these HCT and OCT combinations look like real measured spatiotemporal congested patterns. This visible similarity between the HCT and OCT combinations and real patterns has been used in (Helbing, 2001; Helbing, et al, 1999; Treiber, et al, 2000, 2010; Schönhof and Helbing, 2007, 2009) for the statement that traffic flow models of these authors can describe features of real traffic.

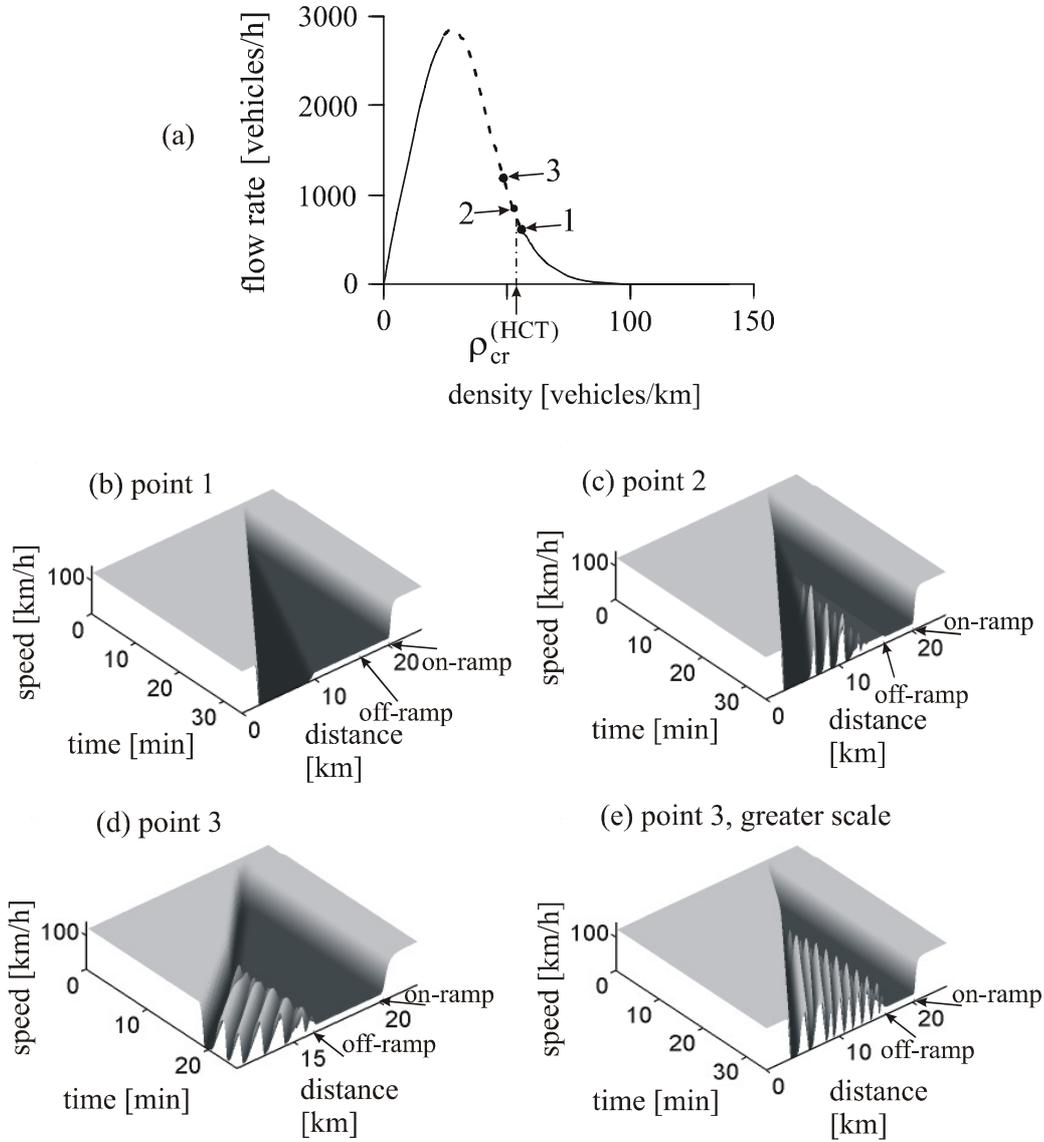

Fig. 15. Simulations of HCT and OCT combinations that occur at downstream on-ramp ( x = 20 km) and upstream off-ramp ( x = 16 km) bottlenecks: (a) Fundamental diagram taken from Fig. 14(a) in which three points 1, 2, and 3 are marked related to different vehicle densities in congested traffic. (b) HCT upstream of the on-ramp bottleneck related to point 1 in (a) for density $\rho_1 > \rho_{cr}^{(HCT)}$ when the flow rate to the off-ramp $q_{off} = 0$. (c) Appearance of OCT upstream of the off-ramp due to instability of HCT upstream of the off-ramp caused by $q_{off} = q_{off\ 1} > 0$ associated with point 2 on the fundamental diagram in (a). (d, e) Stronger growth of OCT upstream of the off-ramp due to increase in the flow rate to the off-ramp to $q_{off\ 2} > q_{off\ 1}$ associated with point 3 on the fundamental diagram in (a); in (d, e) the same pattern in different scales is shown. In (b-e), vehicle speed in space and time. Simulations of Payne-like model.

For example, such HCT and OCT combination occurs in simulations of traffic on a road with two adjacent bottlenecks: downstream on-ramp and upstream off-ramp bottlenecks (Schönhof and Helbing, 2007, 2009).

Firstly, if the on-ramp inflow is great enough and the off-ramp outflow $q_{off}$ is zero, the density in congested traffic $\rho_1$ upstream of the on-ramp can be greater than the critical density $\rho_{cr}^{(HCT)}$ (point 1 in Fig. 15(a)). This results in a HCT model solution upstream of the on-ramp (Fig. 15(b)). As proven in Sect. 5.2.2 above, the HCT has no sense for real traffic.

Now we choose $q_{off} > 0$. Due to the off-ramp outflow $q_{off} > 0$, the flow rate increases and density decreases within this HCT upstream of the off-ramp. As the result, point on the fundamental diagram related to HCT upstream of the off-ramp moves to smaller densities. Specifically, at chosen model parameters point 1 moves to point 2 on the fundamental diagram. However, point 2 is related to the density $\rho_2 < \rho_{cr}^{(HCT)}$, which is on a part of the fundamental diagram for unstable model solutions (a dashed part of the fundamental diagram in Fig. 15(a) left to the critical density $\rho_{cr}^{(HCT)}$). For this reason, the HCT becomes unstable upstream of the off-ramp and transforms into OCT upstream of the off-ramp: HCT and OCT combination occurs spontaneously (Fig. 15(c)).

In a neighborhood of the critical density $\rho_{cr}^{(HCT)}$ for HCT instability, the smaller the density in congested traffic in comparison with $\rho_{cr}^{(HCT)}$, the greater the increment of the growth of fluctuations in the unstable model solutions on the fundamental diagram. As a result, we find that OCT grows quicker upstream of the off-ramp (Fig. 15(d)), when we increase $q_{off}$ and, therefore, point 2 on the fundamental diagram in Fig. 15(a) moves to point 3 associated with a smaller density $\rho_3$ ($\rho_3 < \rho_2$) on the fundamental diagram.

Thus the resulting simulated patterns consist of the HCT solution between the on- and off-ramp bottlenecks and the OCT solution upstream of the off-ramp (Fig. 15(c, d)). Such spatiotemporal combinations of the HCT and OCT model solutions, which look like as GPs (this can be clear seen when we show the pattern in Fig. 15(d) in a greater scale (Fig. 15(e)), have been presented in (Schönhof and Helbing, 2007, 2009; Treiber, et al 2010) as an explanation of measured congested patterns. However, the simulated patterns contain HCT model solutions whose features have no sense for real traffic.

- Thus any explanation of real measured spatiotemporal congested patterns through the use of combinations HCT and OCT model solutions made in (Helbing, 2001; Treiber, et al, 2000, 2010; Schönhof and Helbing, 2007, 2009) is *invalid*.

*5.2.4. Conclusions to criticism of three-phase traffic theory*

(1) Main criticisms of three-phase traffic theory are based on invalid analyses of measured traffic data made in (Treiber, et al, 2000, 2010; Schönhof and Helbing, 2007, 2009), in particular, resulting in the *incorrect* statement about empirical observations of the "boomerang effect", i.e., an F → J transition in real measured traffic data.

(2) Three-phase traffic theory explains real measured spatiotemporal congested patterns through features of SPs and GPs found in this theory (Kerner, 2004a, 2009). Alternative explanations of the same measured congested patterns through the use of diverse combinations of HCT and OCT model solutions made in (Helbing, 2001; Treiber, et al, 2000, 2010; Schönhof and Helbing, 2007, 2009) are considered as additional criticisms of three-phase traffic theory. As proven above, features of HCT model solutions have no sense for real traffic. Therefore, spatiotemporal combinations of the HCT and OCT solutions as well as the associated theoretical diagrams of congested patterns of (Helbing, 2001; Helbing, et al, 1999; Treiber, et al, 2000, 2010; Schönhof and Helbing, 2007, 2009) are not applicable for real traffic flow. In other words, the associated criticisms of three-phase traffic theory are also *incorrect*.

**Acknowledgements:**

I thank Sergey Klenov for discussions and help in numerical simulations.